\documentclass[fleqn,usenatbib]{mnras}

\usepackage{newtxtext,newtxmath}

\usepackage[T1]{fontenc}

\DeclareRobustCommand{\VAN}[3]{#2}
\let\VANthebibliography\thebibliography
\def\thebibliography{\DeclareRobustCommand{\VAN}[3]{##3}\VANthebibliography}

\usepackage{graphicx}	
\usepackage{amsmath}	

\newcommand{\erg}{erg cm$^{-2}$ s$^{-1}$} 
\newcommand{\lum}{erg s$^{-1}$}

\newcommand{\nus}{\emph{NuSTAR}}
\newcommand{\lxp}{\emph{AstroSat}/LAXPC}
\newcommand{\sxt}{\emph{AstroSat}/SXT}
\newcommand{\sax}{SAX J1808.4-3658}
\newcommand{\astrosat}{\emph{AstroSat}}

\title[AstroSat observation of \sax]{The \astrosat\ observation of accreting millisecond X-ray pulsar \sax\ during its 2019 outburst}

\author[R. Sharma et al.]{Rahul Sharma$^{1,2}$\thanks{E-mail: rahul1607kumar@gmail.com; rsharma@rri.res.in},
Andrea Sanna$^{3}$ and
Aru Beri$^{2,4}$
\\
$^{1}$Raman Research Institute, C.V. Raman Avenue, Sadashivanagar, Bengaluru, Karnataka 560080, India\\
$^{2}$Indian Institute of Science Education and Research (IISER) Mohali, Punjab, India 140306\\
$^{3}$Universit\'{a} degli Studi di Cagliari, Dipartimento di Fisica, SP Monserrato-Sestu, KM 0.7, 09042 Monserrato, Italy\\
$^{4}$School of Physics and Astronomy, University of Southampton, Southampton, Hampshire, SO17 1BJ United Kingdom
}

\date{Accepted XXX. Received YYY; in original form ZZZ}

\pubyear{2022}

\begin{document}
\label{firstpage}
\pagerange{\pageref{firstpage}--\pageref{lastpage}}
\maketitle

\begin{abstract}
We report on the analysis of the \astrosat\ dataset of the accreting millisecond X-ray pulsar \sax, obtained during its 2019 outburst. We found coherent pulsations at $\sim 401$ Hz and an orbital solution consistent with previous studies. The 3--20 keV pulse profile can be well fitted with three harmonically related sinusoidal components with background-corrected fractional amplitude of $\sim 3.5 \%$, $\sim 1.2 \%$ and $\sim 0.37 \%$ for fundamental, second and third harmonic, respectively. Our energy-resolved pulse profile evolution study indicate a strong energy dependence. We also observed a soft lag in fundamental and hard lag during its harmonic. The broadband spectrum of \sax\ can be well described with a combination of thermal emission component with $kT \sim 1$ keV, a thermal Comptonization ($\Gamma \sim 1.67$) from the hot corona and broad emission lines due to Fe. 
\end{abstract}




\section{Introduction}

\sax\ was the first X-ray binary to show millisecond X-ray pulsations at $\sim 401$ Hz in a $\sim 2$ hr compact binary orbit \citep{Wijnands1998, Chakrabarty1998}. Since then, 25 such sources with milliseconds pulsations have been discovered \citep[e.g.,][]{DiSalvo2020, Ng2021, Bult2022, Sanna2022b}. These systems are generally referred as Accretion-powered Millisecond X-ray Pulsars (AMXPs) \citep[see e.g.,][for reviews]{Patruno2012, Campana2018, DiSalvo2020}. These systems are transient in nature and observed during outburst phases. AMXPs are a sub-class of low-mass X-ray binaries \citep[LMXBs;][]{Charles2011} where companion's mass is $\lesssim 1M_{\sun}$. 

The magnetic field of AMXPs have been estimated to be of the order of $\sim 10^8-10^9$ G \citep[see e.g.,][Beri et al. 2022 submitted]{Cackett2009, Mukherjee2015, ludlam2017c, Pan2018, Sharma2020}, strong enough to channel at-least part of the accretion stream to the magnetic poles which results in X-ray modulation at Neutron Star (NS) spin. The X-ray spectrum of AMXPs can be well modelled with thermal emission from an accretion disc and/or NS surface, thermal Comptonization and reprocessed emission from the accretion disc in form of reflection spectrum \citep[e.g.,][]{Cackett2009, Cackett2010, Papitto2009, Papitto2010, Papitto2013, Pintore, Salvo2019, Sharma2019}. They mostly show the hard spectral state with some exceptions \citep[][Beri et al. 2022 submitted]{Sharma2019, DiSalvo2020} where they show spectral state transition between the hard and soft spectral states similar to atoll sources \citep[e.g.][]{hasinger}. 

Since 1996, \sax\ has been frequently observed in outburst every 2.5--4 years. A total of ten outbursts (in 1996, 1998, 2000, 2002, 2005, 2008, 2011, 2015, 2019 and 2022) have been observed. 
Over the last two decades, \sax\ has been studied extensively. Aspects of coherent pulsations \citep{Poutanen2003, Jain2007, Hartman2008, Burderi2009, Patruno2017, Sanna2017, Bult2020}, spectral characteristics \citep{Gierlinski2002, Cackett2009, Papitto2009, Salvo2019}, thermonuclear X-ray bursts \citep{intzand1998, Galloway2006, Bhattacharyya2007, Bult2019} and aperiodic and quasi-periodic variability \citep{Wijnands1998b, Wijnands2003, vanStraaten2005, Patruno2009b, Bult2015} have been deeply investigated.
The properties of source have also been deeply investigated during quiescence and outburst in multi-wavelength \citep[e.g., Radio, Optical, UV, etc.][]{Gaensler1999, Wang2001, Wang2013, Cornelisse2009, Heinke2009, Patruno2017, Tudor2017, Baglio2020, Goodwin2020, Ambrosino2021}.

The 2019 outburst of \sax\ started around August 6, 2019 \citep{Bult2019b, Goodwin2019, Parikh2019} which lasted about a month. After episodes of reflaring, source went back into quiescence state
\citep{Baglio2019b, Baglio2019, Bult2019a}, similar to its previous outbursts \citep{Patruno2016}. Twelve days before the on-set of outburst, enhancement in the optical flux was observed \citep{Russell2019b, Goodwin2020}. 
Neutron Star Interior Composition Explorer \citep[\emph{NICER;}][]{Gendreau2017} monitored the 2019 outburst from start to the reflaring states. \citet{Bult2020} performed the coherent timing analysis using NICER observations during the outburst of 2019, and found a secular spin-down of the pulsar at rate of $1.01 (7) \times 10^{-15}$ Hz s$^{-1}$. A very bright Photospheric Radius Expansion (PRE) burst with NICER was also observed \citep{Bult2019}.  
Coherent pulsations in optical and ultraviolet were detected at $\sim 401$ Hz likely to be driven by synchro-curvature radiation in the pulsar magnetosphere or just outside of it \citep{Ambrosino2021}. Recently, \sax\ was again observed in the outburst in August 2022 \citep{Sanna2022ATel}.

\begin{figure}
\centering
 \includegraphics[width=0.9\linewidth]{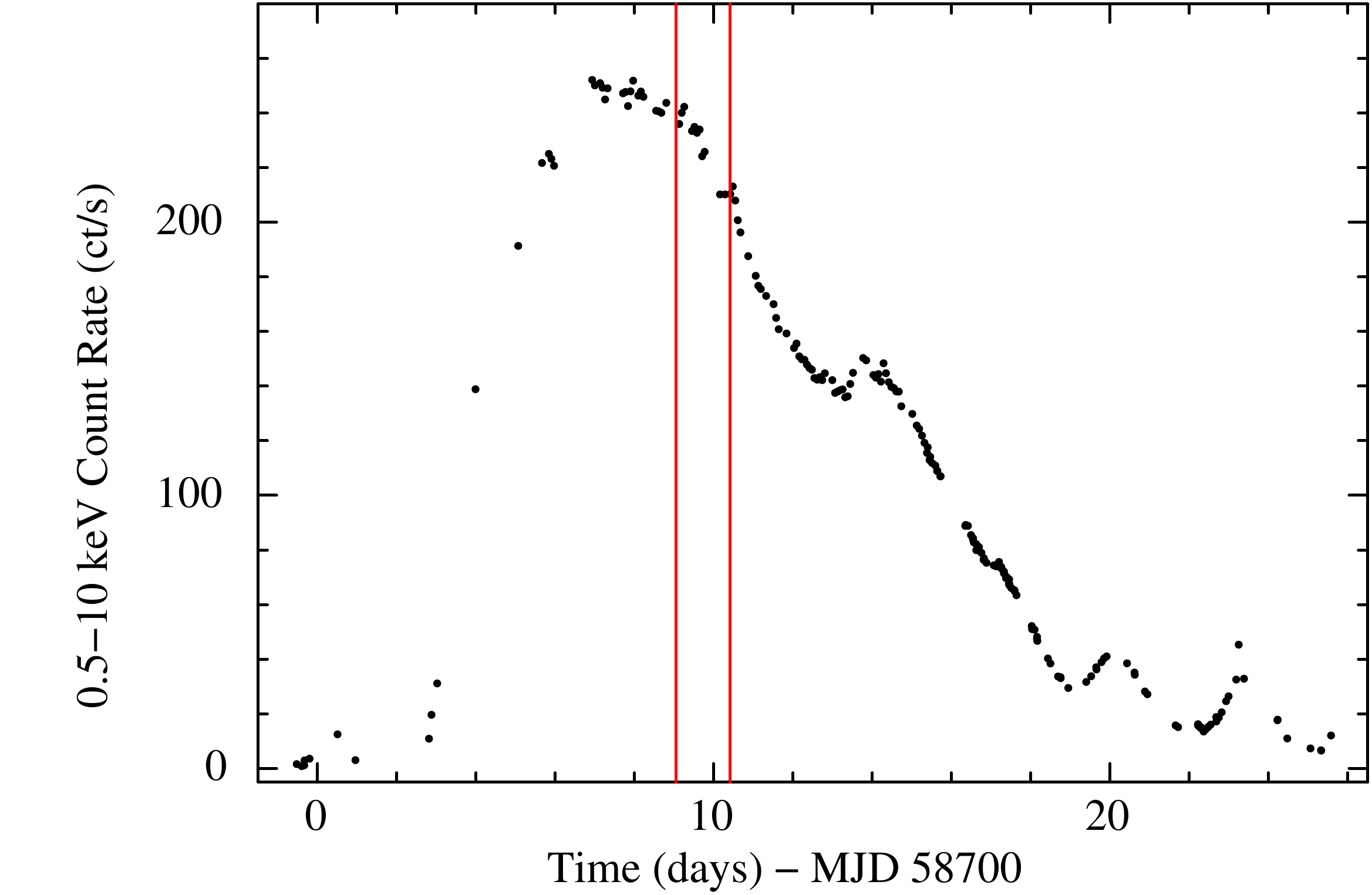}
 \caption{The 0.5--10 keV \emph{NICER} lightcurve of \sax\ during its 2019 outburst. The two vertical red lines represent the time range of the \astrosat\ observation.}
\label{outburst-lc}
\end{figure}

During the 2019 outburst, we also observed \sax\ with \astrosat\ mission under Target of Opportunity (ToO). According to NICER observations, the outburst reached its peak on August 13 (MJD 58708), after which it started to decay \citep{Bult2019a}. The \astrosat\ observation was carried out just after the peak of the outburst (see, Figure \ref{outburst-lc}). In this work, we report for the first time the results from the timing analysis carried out with \lxp\ data and the spectroscopy performed using SXT and LAXPC data.

\section{Observations and data analysis}

\begin{table*}
\caption{Log of X-ray observations.}
\centering
\resizebox{1.9\columnwidth}{!}{
\begin{tabular}{c c c c c c c c}
\hline \hline
Instrument & OBS ID & \multicolumn{2}{c}{Start Time} & \multicolumn{2}{c}{Stop Time} & Mode & Obs span\\
&& (YY-MM-DD HH:MM:SS) & MJD & (YY-MM-DD HH:MM:SS) & MJD & & (ks) \\
\hline
\lxp\ & 9000003090 & 2019-08-14 01:10:46 & 58709.04914 & 2019-08-15 09:56:53 & 58710.41450 & EA & 118 \\
\sxt\ & 9000003090 & 2019-08-14 02:19:02 & 58709.09655 & 2019-08-15 10:30:16 & 58710.43768 & PC & 116\\
\hline
\end{tabular}}
\label{obslog}
\end{table*}

\astrosat\ is India's first dedicated multi-wavelength astronomy satellite \citep{Agrawal2006, Singh2014}, launched in 2015. It has five principal payloads on-board: (i) the Soft X-ray Telescope (SXT), (ii) the Large Area X-ray Proportional Counters (LAXPCs), (iii) the Cadmium-Zinc-Telluride Imager (CZTI), (iv) the Ultra-Violet Imaging Telescope (UVIT), and (v) the Scanning Sky Monitor (SSM). Table~\ref{obslog} gives the log of observations that have been used in this work. We analysed data from SXT and LAXPC only.

\subsection{\lxp}

LAXPC is one of the primary instrument aboard \astrosat. It consists of three co-aligned identical proportional counters (LAXPC10, LAXPC20 and LAXPC30) that work in the energy range of 3--80 keV. Each LAXPC detector independently record the arrival time of each photon with a time resolution of $10 ~\mu$s and has five layers \citep[for details see][]{Yadav2016, Antia2017}.

As LAXPC10 was operating at low gain and detector LAXPC30\footnote{LAXPC30 has been switched off since 8 March 2018 due to abnormal gain changes; see, \url{http://astrosat-ssc.iucaa.in/}} was switched off during the observation, we used only LAXPC20 detector for our analysis. We used the data collected in the Event Analysis (EA) mode and processed using \textsc{LaxpcSoft}\footnote{\url{http://www.tifr.res.in/~astrosat\_laxpc/LaxpcSoft.html}} version 3.1.1 software package to extract light curves and spectra. 
LAXPC detectors have dead-time of $42~\mu$s and the extracted products are dead-time corrected.
The background in LAXPC is estimated from the blank sky observations \citep[see][for details]{Antia2017}. To minimize the background, we have performed all analysis using the data from top layer (L1, L2) of LAXPC20 detector. We have used corresponding response files to obtain channel to energy conversion information while performing energy-resolved analysis.

We corrected the LAXPC photon arrival times to the Solar system barycentre by using the  \textsc{as1bary}\footnote{\url{http://astrosat-ssc.iucaa.in/?q=data\_and\_analysis}} tool. We used the best available position of the source, R.A. (J2000)$=18^h 08^m 27.647^s$ and Dec. (J2000) $=-36^{\circ} 58' 43.90''$ obtained through pulsar timing astrometry \citep{Bult2020}.

\subsection{\sxt}

The Soft X-ray Telescope (SXT) is a focusing X-ray telescope with CCD in the focal plane that can perform X-ray imaging and spectroscopy in the 0.3--7 keV energy range \citep{Singh2016, Singh2017}. 
\sax\ was observed in the Photon Counting (PC) mode with SXT (Table~\ref{obslog}). Level 1 data were processed with \texttt{AS1SXTLevel2-1.4b} pipeline software to produce level 2 clean event files. The level 2 cleaned files from each orbits were merged using SXT Event Merger Tool (Julia Code). This merged event file was then used to extract image, light curves and spectra using the ftool task \textsc{xselect}, provided as part of \textsc{heasoft} version 6.29. 
A circular region with radius of 15 arcmin centered on the source was used. For spectral analysis, we have used the blank sky SXT spectrum as background (SkyBkg\_comb\_EL3p5\_Cl\_Rd16p0\_v01.pha) and spectral redistribution matrix file (sxt\_pc\_mat\_g0to12.rmf) provided by the SXT team\footnote{\url{http://www.tifr.res.in/~astrosat\_sxt/dataanalysis.html}}. We generated the correct off-axis auxiliary response files (ARF) using sxtARFModule tool from the on-axis ARF (sxt\_pc\_excl00\_v04\_20190608.arf) provided by the SXT instrument team.


\begin{figure}
\centering
\includegraphics[width=\linewidth]{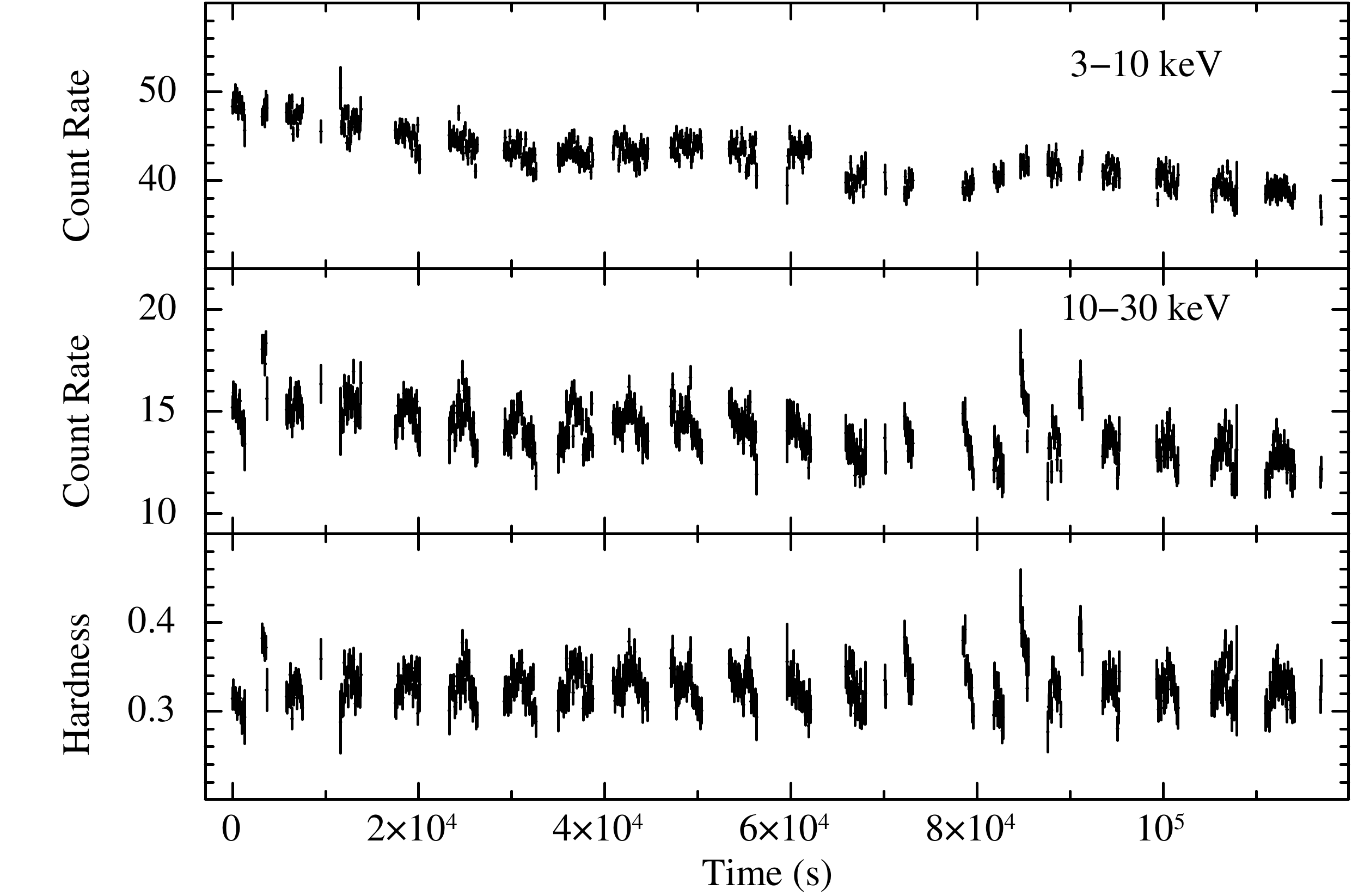}
 \caption{The top two panels show background corrected light curve from LAXPC20 binned at 100 sec in energy range of 3--10 keV and 10--30 keV, respectively. The bottom shows the hardness ratio (=10--30 keV/3--10 keV) binned at 100 sec.}
\label{hr}
\end{figure}

\section{Results}

\subsection{Light Curve}

Fig. \ref{outburst-lc} shows the light curve of \sax\ during its 2019 outburst as observed with NICER. The \astrosat\ observation (red vertical lines) was carried out near the peak of the outburst. Fig. \ref{hr} shows the background corrected light curve extracted from LAXPC20 binned at 100 sec in two different energy bands 3--10 keV (top panel) and 10--30 keV (middle panel). The LAXPC light curves show the persistent emission separated by data gaps due to Earth occultation and South Atlantic Anomaly (SAA) passage. No type-I X-ray burst was observed during the observation. The bottom panel presents the hardness ratio calculated using light curves in the two energy bands 3--10 keV and 10--30 keV. As hardness ratio is observed to be constant during the observation, this suggests that source did not seem to change spectral state within the observation duration.

\subsection{Aperiodic Timing}

We created the power density spectrum (PDS) using LAXPC20 event data-set in the 3--20 keV energy range. We binned the data at 0.5 ms to have Nyquist frequency of 1000 Hz and used $\sim 262$ sec segments to calculate the Fourier transform. All power spectra were averaged and rebinned geometrically with factor of 1.05. No background correction was applied. The power spectra was calculated using rms normalization and Poisson noise was subtracted using \textsc{ftool} \texttt{powspec norm=-2}. The PDS of X-ray binaries can be described in terms of a sum of Lorentzian functions \citep[e.g., ][]{Nowak2000, Belloni2002}. The Lorentzian profile is a function of frequency and can be defined as
   \begin{eqnarray}
   \label{eqn:lore}
   P(\nu) = \frac{r^2 \Delta}{2 \pi} \frac{1}{(\nu - \nu_0)^2 + (\Delta/2)^2},
   \end{eqnarray}
where $\nu_0$ is the centroid frequency, $\Delta$ is the full-width at half-maximum (FWHM), and $r$ is the integrated fractional rms. The quality factor of Lorentzian $Q=\nu_0/\Delta$ used to differentiate if a feature is a Quasi-Periodic Oscillations (QPO) or noise. The components with $Q>2$ are generally considered as QPOs, otherwise band-limited noise \citep[e.g.,][]{Belloni2002, vanderKlis2004, Bult2015}. 

\begin{figure}
\centering
\includegraphics[width=0.9\columnwidth]{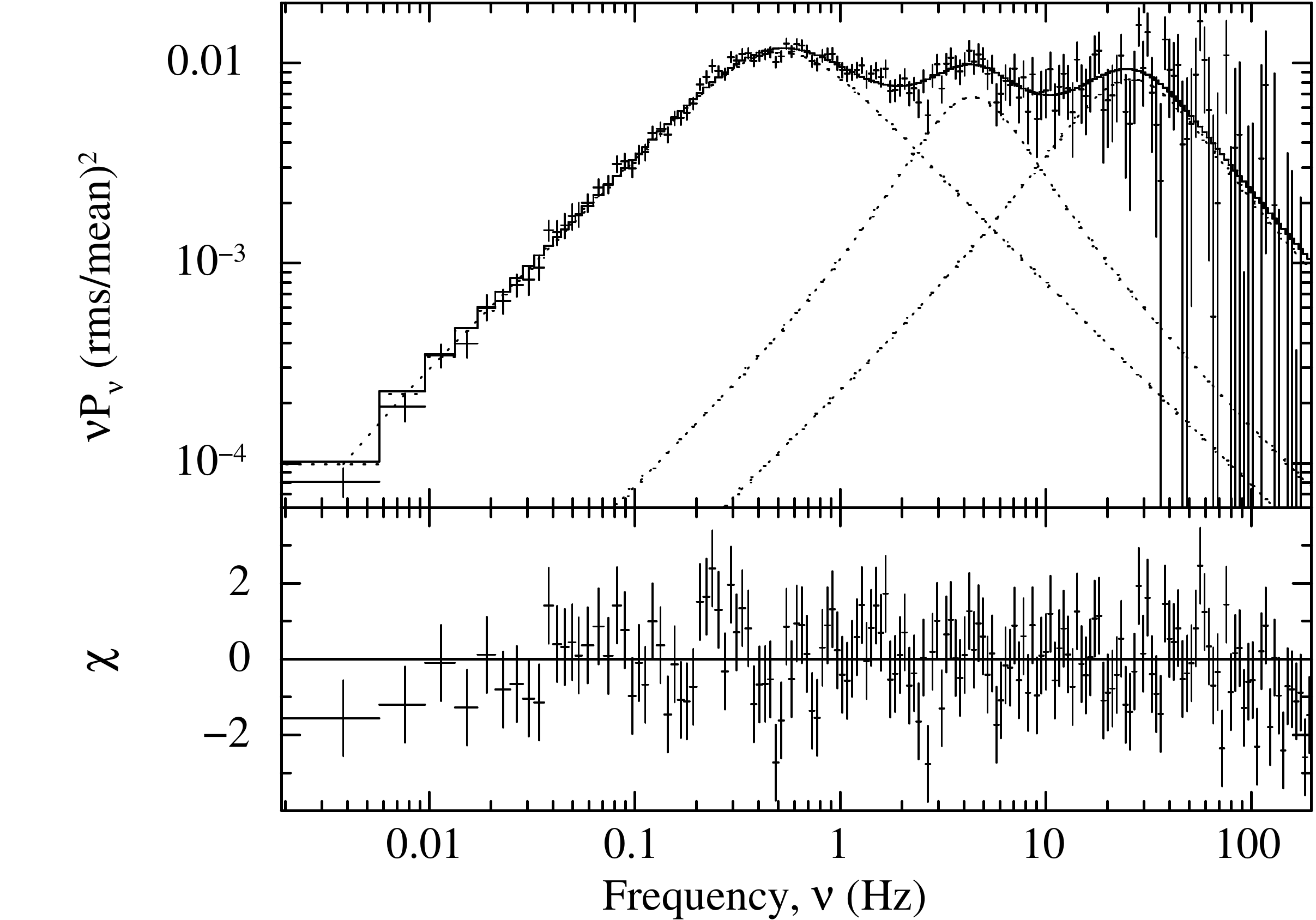}
\caption{Power density spectrum of \sax\ in the 3--20 keV energy range. The PDS is rms normalized and white noise subtracted. Three Lorentzian functions are used to model the PDS in 0.004--200 Hz.}
\label{fig:pds}
\end{figure}

\begin{table}
\centering
\caption{The best-fit of the PDS. Reported errors are at $1\sigma$ confidence level for single parameter.}
\resizebox{0.9\linewidth}{!}{
\begin{tabular}{c c c c}
\hline
Component &  1 & 2 & 3 \\
\hline
Freq., $\nu_0$ (Hz) & $0.1675_{-0.016}^{+0.015}$  & $3.33_{-0.74}^{+0.37}$ & $17.1_{-6.3}^{+5.6}$ \\[1ex]
FWHM, $\Delta$ (Hz) & $0.96 \pm 0.05$  & $5.7_{-1.5}^{+2.1}$ & $39.6_{-6.4}^{+7.2}$ \\[1ex]
Ch. Freq., $\nu_{\rm max}$ (Hz) & $0.508$ & $4.38$ & $26.16$ \\[1ex]
rms (\%) & $17.5^{+0.3}_{-0.4}$  & $11.0_{-1.4}^{+1.8}$ & $13.2 \pm 1.2$ \\[1ex]
            
$\chi^2/{\rm dof}$ &  & 175/151 & \\
\hline
\end{tabular}}
\label{table:pds}
\end{table}

\sax\ is known to show peaked noise, low-frequency QPOs and kHz QPOs \citep[e.g.,][]{Wijnands1998b, Wijnands2003, vanStraaten2005, Patruno2009b, Bult2015}. 
Fig. \ref{fig:pds} shows the PDS of \sax\ from LAXPC20 in 0.004--200 Hz, as no significant power is observed above 200 Hz. It shows only the red noise and no low-frequency QPO was observed. The PDS can be well modelled with combination of three Lorentzian functions (see, Table \ref{table:pds}). Generally, each Lorentzian of the noise component can be expressed with characteristic frequency $\nu_{\rm max} = \sqrt{\nu_0^2 + (\Delta/2)^2}$ and fractional rms amplitude \citep{Belloni2002, Bult2015}. We found the $\nu_{\rm max}$ of $\sim$ 0.5, 4.4, 26 Hz and rms of $\sim$ 17.5, 11, 13 \% of the three Lorentzians, respectively. 

\subsection{Coherent Timing analysis}

\begin{figure}
\centering
\includegraphics[width=\columnwidth]{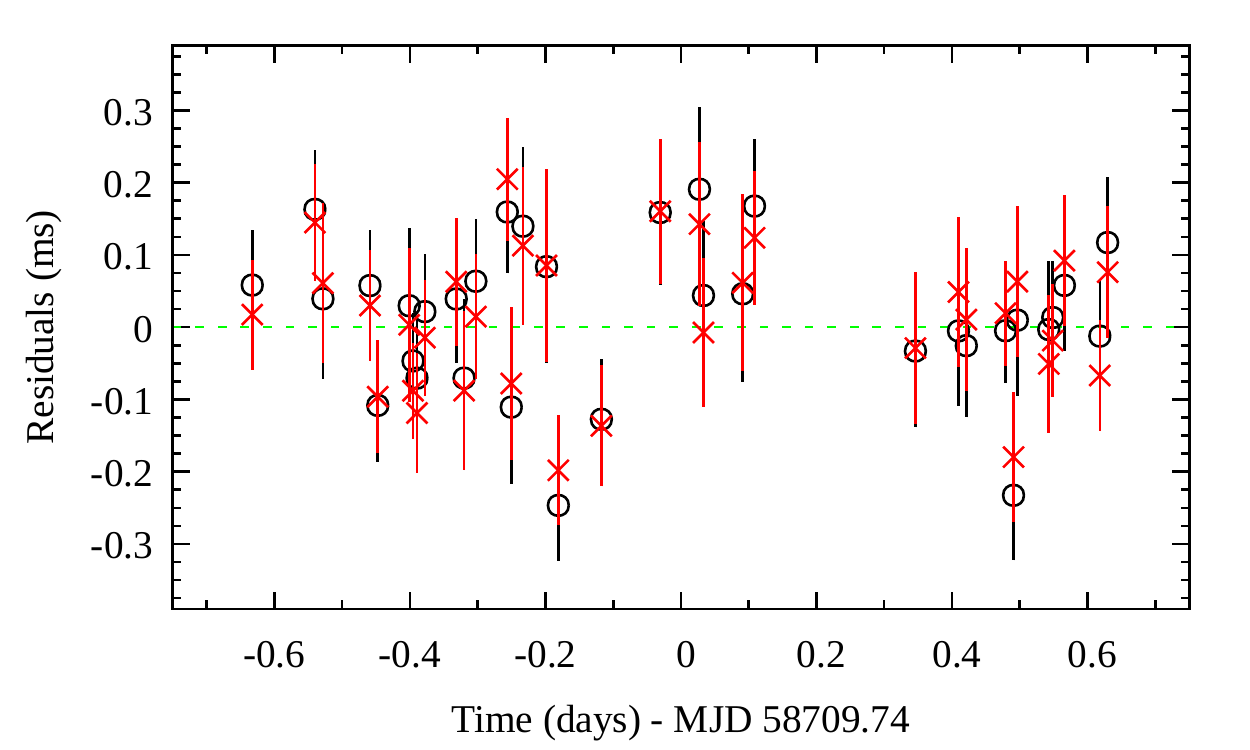}
\caption{The residuals obtained from fitting the pulse phase delays of the fundamental component. Red cross points represents the residuals with equation \ref{eqn:quad} (quadratic+orbital) and black circles represents the residuals with only quadratic function ($R_{\rm orb} (t) = 0$).}
\label{fig:phase}
\end{figure}

The pulses from rotating NS can lose coherence within a relatively short timescale  
due to Doppler modulations of arrival time of pulses in short orbital period of $\sim 2$ hr. In addition, the small pulse fraction can also makes it difficult to detect pulse arrival times from small segments of the light curve. Therefore, to detect the pulsations or true spin frequency during these observations, the light curve should be corrected for the binary motion. So, we corrected the photon time of arrivals of the \lxp\ data-set for the delays caused by the binary motion. The delay induced by the orbital motion can be written as \citep{Burderi2007}:

\begin{eqnarray} 
\label{eq:bary}
\frac{z(t)}{c}= \frac{a_x \sin i}{c}\,\sin\Big(\frac{2\pi}{P_{\rm orb}} \,(t-T^\star)\Big),
\end{eqnarray}
where $a_x \sin{\textit{i}/c}$ is the projected semi-major axis of the NS orbit in light seconds, $P_{\rm orb}$ is the orbital period, and $T^\star$ is the time of passage from the ascending node.

\citet{Bult2020} found the latest updated orbital solution at the outburst of 2019 using NICER observations. Therefore, we corrected photon time of arrivals adopting the 2019 orbital ephemeris reported in \citet{Bult2020}. 
We then applied epoch-folding techniques to search for X-ray pulsation around the spin frequency value $\nu_0$ = 400.97520983 Hz, corresponding to the spin frequency measured during the same outburst with NICER \citep{Bult2020}. We used 16 phase bins to sample the signal. We explored the frequency space with steps of $10^{-8}$ Hz for a total of 10001 steps. The most significant pulse profile was been obtained for an average local spin frequency of $\bar{\nu}=400.97520994$ Hz. 

We also investigated the evolution of the pulse phase delays. We divided the data in time intervals of $\sim 500$ sec and epoch-folded each segment in $8$ phase bins at the mean spin frequency $\bar{\nu}$ with respect to the epoch $T_0 = 58709.74$ MJD. We modelled each pulse profile with a sinusoidal function for the fundamental component to determine the corresponding sinusoidal amplitude and the phase delay. We selected only folded profiles with ratio between the sinusoidal amplitude and the corresponding 1 sigma error larger than 3. The fractional amplitude (non-background corrected) of the fundamental varies between $\sim 2.2-3.7$\%.

We modelled the time evolution of the pulse phase delays obtained from the fundamental component with the following model \citep[see e.g.][]{Burderi2007, Sanna2020, Sanna2022}:
\begin{eqnarray}
\label{eqn:quad}
\Delta \phi (t) = \phi_0 + \Delta \nu_0 (t-T_0) - \frac{1}{2} \dot{\nu} (t - T_0)^2  + R_{\rm orb}(t),
\end{eqnarray}
where $\Delta \nu_0 = (\nu_0 - \bar{\nu})$ is spin frequency correction and $\dot{\nu}$ is the spin frequency derivative, estimated with respect to the reference epoch $T_0$. $R_{\rm orb} (t)$ models the differential correction to the ephemeris used to generate the pulse phase delays \citep[see e.g.,][]{Deeter1981}. As the orbital solution is already updated at the 2019 outburst, we set $R_{\rm orb} (t) = 0$ (case 1) and measured the updated spin frequency and its derivative. As a second case, we also allowed $R_{\rm orb} (t)$ and measured the orbital parameters along with spin frequency and its derivative. Best-fit parameters are reported in Table \ref{timing-log}. The residuals obtained after the best-fit subtracted from phase delays are shown in Fig. \ref{fig:phase} where black points represents quadratic only, while red points quadratic+orbital.
The obtained best-fit frequency is well consistent with spin frequency observed with NICER \citep{Bult2020}. We obtained a $3\sigma$ upper limits on the spin frequency derivative of $-4 \times 10^{-11}$ Hz s$^{-1}$ to $6.2 \times 10^{-11}$ Hz s$^{-1}$ for case 1 and $-3.9 \times 10^{-11}$ Hz s$^{-1}$ to $7.5 \times 10^{-11}$ Hz s$^{-1}$ for case 2. The obtained orbital solution is also well consistent with \citet{Bult2020}'s solution within the errors. 

\begin{figure}
\centering
\includegraphics[width=\columnwidth]{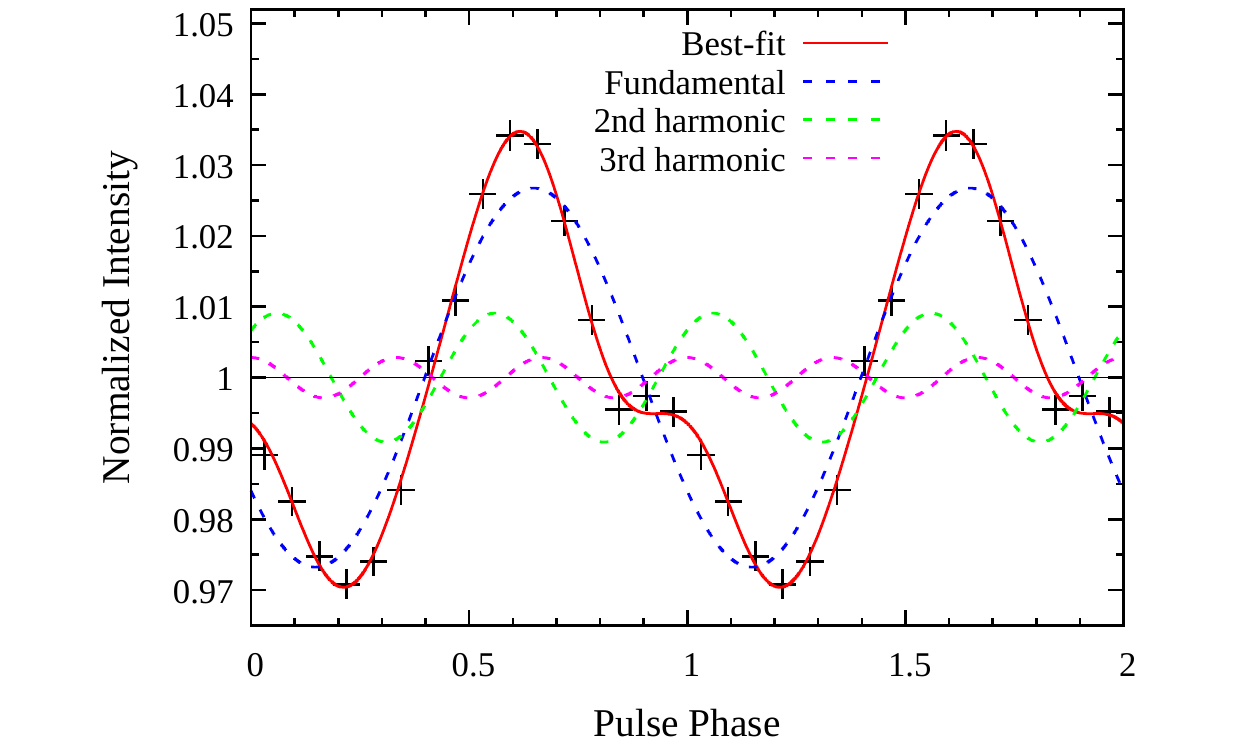}
 \caption{The average pulse profile (black points) of \sax\ obtained epoch-folding the \lxp\ data with $\bar{\nu}$ in the energy range 3--20 keV after correcting for the orbital solution. The epoch used was $T_0=58709.74$ MJD. The best-fitting model (red solid line) is the superposition of three sinusoidal functions with harmonically related periods. The fundamental, second and third harmonic components are presented by the blue, green and magenta dashed lines, respectively. For clarity, we show two cycles of the pulse profile.}
\label{fig:pp}
\end{figure}

Fig.~\ref{fig:pp} presents the best pulse profile obtained by epoch-folding the LAXPC dataset at $\bar{\nu}$ and sampling the signal in 16 phase bins in the energy range of 3--20 keV. The pulse shape is well fitted with three harmonically related sinusoidal function with background corrected fractional amplitude of $3.54 (7) \%$, $1.21 (7) \%$ and $0.37 (7) \%$ for the fundamental, second and third harmonic, respectively. 

\begin{figure}
\centering
\includegraphics[width=0.9\columnwidth]{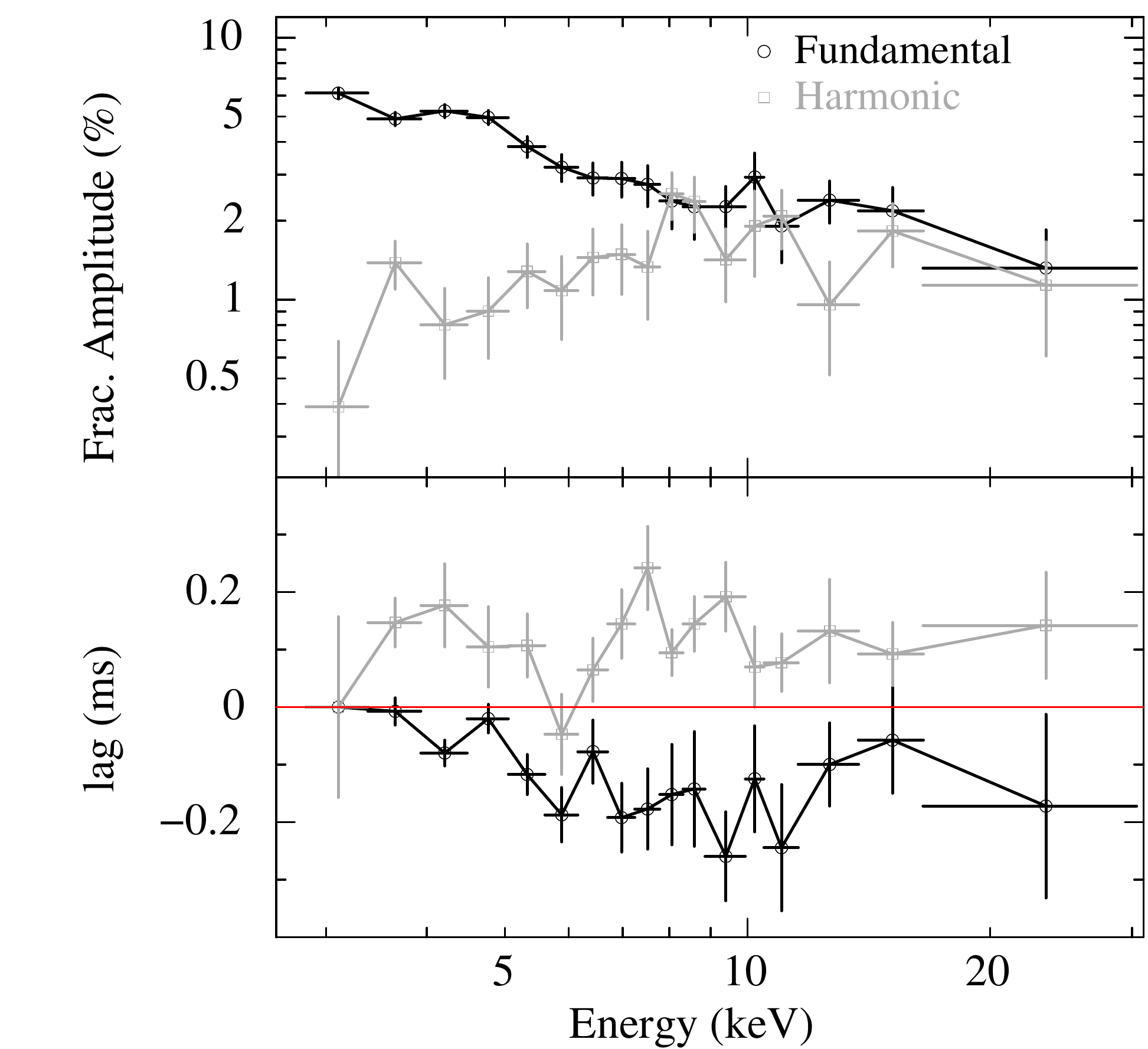}
 \caption{\textit{Top panel:} The energy dependence of pulse fractional amplitude for the fundamental and harmonic. \textit{Bottom panel:} The observed time lag of the fundamental and harmonic component of the pulse with energy. The zero lag corresponds to the phase of the pulse profile in 2.83--3.38 keV (channel no. 5 of LAXPC20 corresponding to 3 keV).}
\label{fig:amp}
\end{figure}

We also checked for the energy dependence of the pulse amplitude and phase. We divided the whole dataset depending on the different energy range and epoch-folded each with $\bar{\nu}$. Some of the energy-resolved pulse profiles are presented in fig. \ref{en-profile}.
Each folded pulse profile was then fitted with two harmonically related sinusoidal functions (third harmonic was not detected in energy-resolved pulse profiles, likely due to poor statistics of selected intervals). Fig. \ref{fig:amp} (top panel) shows the variation of background-corrected fractional amplitude of the fundamental (black points) and harmonic (grey points) components of the pulse. The fractional amplitude of the fundamental was found to be $\sim5$\% in 3--5 keV energy range and above 5 keV a constant decrease was observed with energy. While the harmonic component increased with energy. Fig. \ref{fig:amp} (bottom panel) shows the time lag (calculated from the phase lag) of the fundamental and harmonic component of pulse with energy. The zero lag value corresponds to the phase of pulse profile in the energy band of $2.83-3.38$ keV. It shows negative time lag  (i.e. high-energy emission is coming earlier than the softer emission) of 0.1--0.2 ms for fundamental component at higher energies (above 5 keV) while, the harmonic component shows the positive lag of similar order. By comparing top and bottom plots of Fig. \ref{fig:amp}, the phase lags seems to become specular as soon as the fractional amplitude of the two components reaches similar values (above $\sim 7$ keV). 

\begin{table}
\caption{Timing solution of \sax\ during the outburst of 2019. Errors are at $1 \sigma$ confidence level.}
\centering
\resizebox{1.05\columnwidth}{!}{
\begin{tabular}{c c c}
\hline \hline
Parameter & Case 1 & Case 2 \\
\hline
Spin Frequency, $\nu$ (Hz) & 400.97521014(21) & 400.97521004(26)\\[1ex]
Spin Frequency derivative, $\dot{\nu}$ (Hz/s)& $1.1(1.7) \times 10^{-11}$ & $1.8(1.9) \times 10^{-11}$ \\[1ex]
Ascending node passage, $T^{\star}$ (MJD) & 58715.0221031$^a$ & 58715.02212 (7) \\[1ex]
Orbital period, $P_{\rm orb}$ (s) & 7249.1552$^a$ & 7249.16 (10)\\[1ex]
Projected semi-major axis, $a_x \sin i/c$ (lt-ms) & 62.8091$^a$ & 62.83 (3)\\[1ex]
Eccentricity, $e$ & \multicolumn{2}{|c|}{$0^a$}  \\[1ex]
Reference epoch, $T_0$ (MJD) & \multicolumn{2}{|c|}{58709.74}  \\[1ex]
$\chi^2$/d.o.f. & 46/31 & 41/28\\
\hline
\multicolumn{3}{l}{$^a$taken from \citep{Bult2020}}.\\
\end{tabular}}
\label{timing-log}
\end{table}


\begin{table}
\centering
\caption{The best-fit spectral parameters obtained from the SXT+LAXPC spectrum of \sax. Reported errors and limits are at 90\% confidence level for single parameter.}
\resizebox{0.8\linewidth}{!}{
\begin{tabular}{c c c}
\hline
Model & Parameters & SXT+LAXPC\\
\hline
TBabs & $N_H$ ($10^{22}$ cm$^{-2}$) & $0.056 \pm 0.007$ \\
\\
DiskBB & $kT_{\rm in}$ (keV) & $1.08^{+0.03}_{-0.02}$ \\[1ex]
       & Norm & $7.8^{+1.2}_{-1.0}$ \\
\\      
NTHcomp & $\Gamma$ & $1.67^{+0.02}_{-0.01} $ \\[1ex]
        & $kT_{\rm seed}$ (keV) & $0.322 \pm 0.025$ \\[1ex]
        & \textit{inp\_type} & 0 \\[1ex]
        & $kT_e$ (keV) & $100^{\rm fixed}$ \\[1ex]
        & Norm ($10^{-2}$)  & $3.7 \pm 0.3$ \\
\\
Gaussian 1 & $E_{\rm Line}$ (keV) & $7.0^{\rm pegged}_{-0.25}$ \\[1ex]
            & $\sigma$ (keV) & $1.0^{\rm pegged}_{-0.2}$ \\[1ex]
            & Eqw (keV) & $0.26^{+0.11}_{-0.14}$ \\[1ex]
            & Norm ($10^{-4}$)  & $7 \pm 2$ \\[1ex]
\\
Gaussian 2 & $E_{\rm Line}$ (keV) & $0.86^{+0.04}_{-0.07}$ \\[1ex]
            & $\sigma$ (keV) & $0.08^{+0.05}_{-0.03}$ \\[1ex]
            & Eqw (keV) & $0.022 \pm 0.010$ \\[1ex]
            & Norm ($10^{-3}$)  & $2.0^{+1.4}_{-0.8}$ \\[1ex]
\\            
Constant  & $C_{\rm LAXPC}$ & 1 (fixed) \\[1ex]
          & $C_{\rm SXT}$ & $1.04 \pm 0.02$ \\
\\
SXT Gain & Gain (eV) & $34^{+2}_{-4}$ \\
\\

Unabs. Flux$^{*}$ & $F_{0.1-100~\rm keV}$  & $1.3 \times 10^{-9}$ \\

Luminosity$^{\dagger}$ & $L_X$ (\lum) & $1.9 \times 10^{36}$ \\[1ex]
\\            
          & $\chi^2/{\rm dof}$ & 700/576 \\
\hline
\multicolumn{3}{l}{$^*$ 0.1--100 keV Unabsorbed flux in units of \erg.}\\
\multicolumn{3}{l}{$^{\dagger}$X-ray Luminosity in 0.1--100 keV for a distance of 3.5 kpc.}\\

\end{tabular}}
\label{spec}
\end{table}


\subsection{Broadband Spectral Analysis}

\begin{figure}
\centering
\includegraphics[width=\columnwidth]{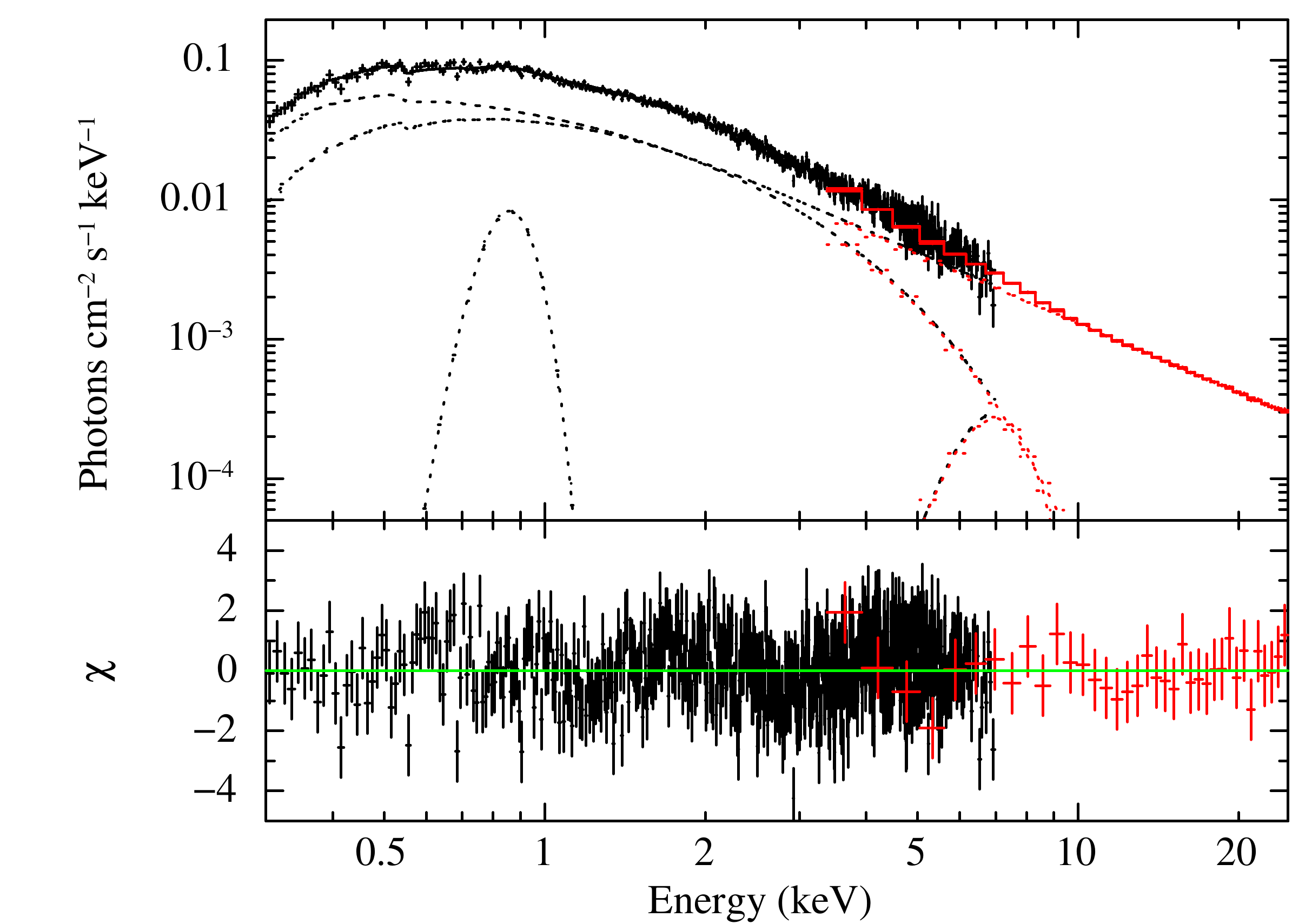}
 \caption{Broadband spectrum of \sax\ from SXT and LAXPC instrument fitted simultaneously with best fit model \texttt{tbabs*(diskbb+nthcomp+gaussian+gaussian)}. Panel below show residuals ($\chi$=(data-model)/error) with respect to best fit model.} 
\label{fig:spectra}
\end{figure}

We modelled the SXT (0.3--7 keV) and LAXPC (3--25 keV\footnote{Above 25 keV, LAXPC20 showed systematic residuals \citep{Sharma2023}.}) spectrum simultaneously. Both spectra were grouped using \textsc{grppha} to have a minimum count of 20 counts per bin. An inter-instrumental calibration constant was added, which was fixed to 1 for LAXPC and allowed to vary for SXT. We also allowed the gain of response matrix for SXT to vary, with slope fixed at 1. A gain offset of $\sim 34$ eV was obtained. A systematic uncertainty of 2\% was also introduced during the spectral fitting \citep{Antia2017, Sharma2020}. We used \textsc{xspec} \citep{Arnaud} for the spectral fitting with the component \texttt{tbabs} to model interstellar neutral hydrogen absorption \citep{Wilms}.

To model the broadband spectrum, we first started with the absorbed thermal Comptonized emission model, \texttt{nthcomp} \citep{Zdziarski, Zycki}. The \texttt{nthcomp} model pictures that the hot electrons Compton up-scatter the seed photons originating from the NS surface/boundary layer or inner accretion disc. Using parameter \textit{inp\_type}, input seed type can be selected. During all fitting, the electron temperature was unconstrained so we fixed it at reasonable value of 100 keV. The obtained fit with \texttt{tbabs$\times$nthcomp} was unacceptable with either of input seed photon types ($\chi^2/{\rm dof} \sim 1133/583$). 

We then added a thermal component in form of a single temperature blackbody. It did not provide a satisfactory fit ($\chi^2/{\rm dof} = 826/581$). We then replaced single temperature blackbody with multi-coloured blackbody \citep[\texttt{diskbb};][]{Mitsuda} as the thermal component and set blackbody as seed (\textit{inp\_type} = 0). It provided slight improvement with $\chi^2/{\rm dof} = 758/581$. The fit showed systematic residuals around 7 keV which hints at the Fe emission line \citep[e.g., ][]{Cackett2009, Papitto2009, Salvo2019}, so we added a Gaussian component to the model. During the fit, we constrained the emission line energy to be within 6.4--7.0 keV energy range (corresponds to Fe K emission). We found that the Gaussian line energy pegged at upper limit of 7 keV with width broader than $1$ keV. So we constrained the Gaussian line width to be within 0 to 1 keV. Addition of this Gaussian component improved the fit significantly to $\chi^2/{\rm dof} \sim 726.7/579$ with F-test probability of improvement by chance $\sim 5 \times 10^{-6}$. 
The equivalent width of this iron emission line feature was about $\sim 0.26$ keV, while $\sim 0.1$ keV was observed with the \nus\ observation of 2015 outburst \citep{Salvo2019}. Further, the obtained spectral residuals at low energies hints at some emission feature around 0.9 keV, possibly due to Ne \textsc{ix} or Fe L blend. Similar emission features have been seen with XMM-Newton observation \citep{Salvo2019}. This encouraged us to add another Gaussian emission line model. Addition of emission line at $\sim 0.85$ keV further improved the fit to $\chi^2/{\rm dof} \sim 700/576$ with F-test probability of improvement by chance $\sim 9 \times 10^{-5}$. The best fit 0.5--25 keV persistent spectrum of \sax\ is shown in Fig. \ref{fig:spectra} and the best fit parameters are listed in Table \ref{spec}. We also estimated the 0.1--100 keV unabsorbed flux of $1.3 \times 10^{-9}$ \erg\ from the best fit model which translate to X-ray luminosity of $1.9 \times 10^{36}$ \lum\ for a distance of 3.5 kpc \citep{Galloway2006}. We also tried to change the input seed source to disc and we obtained similar spectral fit parameters except $kT_{\rm seed} = 0.57$ keV with $\Delta \chi^2 = +6$ for no additional dof.



\section{Discussion and Conclusion}

\sax\ was observed in its ninth outburst in 2019. The \astrosat\ observed it for $\sim 1.5$ days near the peak of the outburst. We report results from our broadband timing and spectral study of \sax\ performed using \astrosat\ data. The 0.004--200 Hz PDS showed noise components which can be described by three Lorentzian functions with integrated rms of 17.5\%, 11\% and 13\%. 
Moreover, we found coherent X-ray pulsations at $\sim401$ Hz. Pulse timing results obtained are compatible within the errors with the solution obtained with NICER \citep{Bult2020}. The pulse profiles obtained in the energy bands 3--20 keV suggest a fractional amplitude of 3.5\%, 1.2\% and 0.37\% for fundamental, second and third harmonic, respectively. 
Using pulse phase delays, we obtained an upper limits ($3\sigma$) on the spin down rate of $\sim -4 \times 10^{-11}$ Hz s$^{-1}$ and spin up rate of $\sim 7 \times 10^{-11}$ Hz s$^{-1}$ for the \astrosat\ observation duration.

We also checked for the dependence of fractional amplitude and phase lag on the energy in the 3--30 keV range. The fractional amplitude and phase lag were found to be energy dependent as previously observed \citep[e.g.,][]{Hartman2009, Sanna2017, Bult2020}. 
The fundamental component showed fractional amplitude of $\sim 5\%$ up to 5 keV, after that it starts decreasing and reached around 2\%. While the harmonic component showed an increasing trend in 3--30 keV energy range from $\sim 1 \%$ to 2\%. Similar trend of increase in the fractional amplitude of the harmonic with energy has been observed in previous outbursts. The energy dependence of the fractional amplitudes varies considerably between different outbursts \citep[e.g.,][]{Hartman2009, Patruno2009, Sanna2017}. Also, we did not detect any drop in fractional amplitude in 6--7 keV energy range (correspondence to the Fe line) as observed in the 2008 and 2015 outburst \citep{Patruno2009, Sanna2017}.

\sax\ is known to show energy dependent time lags in which soft X-rays lag hard X-rays, i.e. the pulse peaks at a later phase in softer energies \citep[e.g.,][]{Cui1998, Gierlinski2002, Hartman2009, Ibragimov2009}. We also found similar trend with LAXPC. The lags for fundamental component is soft and increased from 3 to $\sim 10$ keV. While the harmonic component showed opposite trend, i.e. hard lags of similar order. The lags of \sax\ are flux dependent and increases with drop in accretion rate. The soft lags in the fundamental observed throughout the outbursts while lags for harmonic are less pronounced or becomes hard lags when source flux is high \citep{Hartman2009, Ibragimov2009}.  
These lags can be explained by the model where the soft thermal component and Comptonization emissivity (or beaming) patterns are different, which is affected in a different way by the fast stellar rotation \citep{Gierlinski2002, Poutanen2003, Ibragimov2009}. 

The 0.3--25 keV broadband spectrum of \sax\ was well described by the combination of a thermal component as multi-coloured blackbody, Comptonized emission of soft thermal photons and emission lines. A multi-coloured disc-like component statistically better fits the soft continuum than a single temperature blackbody. The thermal emission was found to be of $\sim 1$ keV temperature with compact emission size, possibly from NS or accretion column. The Comptonized emission associated to a hot corona is characterized by a photon index of $1.67$. A power-law photon index of 1.89 was measured from \nus\ observation of 2019 at the outburst peak \citep{Sanna2019}. The electron temperature of corona was unconstrained and input seed temperature was found to be $\sim 0.3$ keV for blackbody seed and $\sim 0.5$ keV for diskbb seed. This soft component provides the seed photons to the Comptonized medium which is not directly visible.  
The best-fitting spectral parameters suggest that the source was in the hard spectral state during the \astrosat\ observation (Table \ref{spec}). Throughout the outburst, the X-ray spectrum was hard during both the main outburst peak and the reflaring period \citep{Bult2019b, Bult2019a, Sanna2019}. The hardness ratio observed from \astrosat\ did not show any significant variations, suggesting source to be in the same state throughout the observation duration. Also, no significant variations in the hardness ratio was observed with \emph{Swift}/XRT during the whole outburst of 2019 \citep{Baglio2020}. Therefore, 
\sax\ never left the hard state and went directly from a hard-state main outburst to the reflaring state \citep[see e.g.][]{Patruno2016, Baglio2020}. During the outburst of 2015, \citet{Salvo2019} reported the state transition in \sax\ with XMM-Newton (soft state) and \nus\ (hard state) observations.

Generally, AMXPs show the hard spectrum with electron temperature of few tens of keV \citep{Falanga2005, Gierlinski2005, Papitto2010, Papitto2013, Wilkinson2011, Sanna2018a, Sanna2018b, Salvo2019}. Another AMXP SAX J1748.9-2021 also showed the spectral state transition from hard to soft state during bright outbursts \citep{Patruno2009a, Pintore, Li2018, Sharma2019, Sharma2020}. However, during the fainter outburst where SAX J1748.9-2021 have similar luminosities as of \sax\ ($\sim 10^{36}-10^{37}$ erg s$^{-1}$), source seems to stay in the hard state and do not show state transitions \citep{intzand1999, Sharma2020}. 
So AMXPs show hard spectrum like faint transients when emitting at low luminosities. 

\section*{Acknowledgements}

RS would like to thank the University of Cagliari for the hospitality during the research visit, where this work was started. The research visit was funded by the Committee on Space Research (COSPAR) and Indian Space Research Organisation (ISRO). RS and AB are supported by an INSPIRE Faculty grant (DST/INSPIRE/04/2018/001265) awarded to AB by the Department of Science and Technology, Govt. of India and also acknowledges the financial support of ISRO under AstroSat archival Data utilization program (No.DS-2B-13013(2)/4/2019-Sec. 2). AB is grateful to both the Royal Society, U.K and to SERB (Science and Engineering Research Board), India. This research has made use of the \astrosat\ data, obtained from the Indian Space Science Data Centre (ISSDC). We thank the LAXPC Payload Operation Center (POC) and the SXT POC at TIFR, Mumbai for providing necessary software tools. We also thank the anonymous referee for the valuable comments on the manuscript.


\section*{Data Availability}

Data used in this work can be accessed through the Indian Space Science Data Center (ISSDC) at 
\url{https://astrobrowse.issdc.gov.in/astro\_archive/archive/Home.jsp}.



\bibliographystyle{mnras}
\bibliography{sample} 




\appendix

\section{Energy-resolved pulse profiles}

\begin{figure}
\centering
\includegraphics[width=0.9\linewidth]{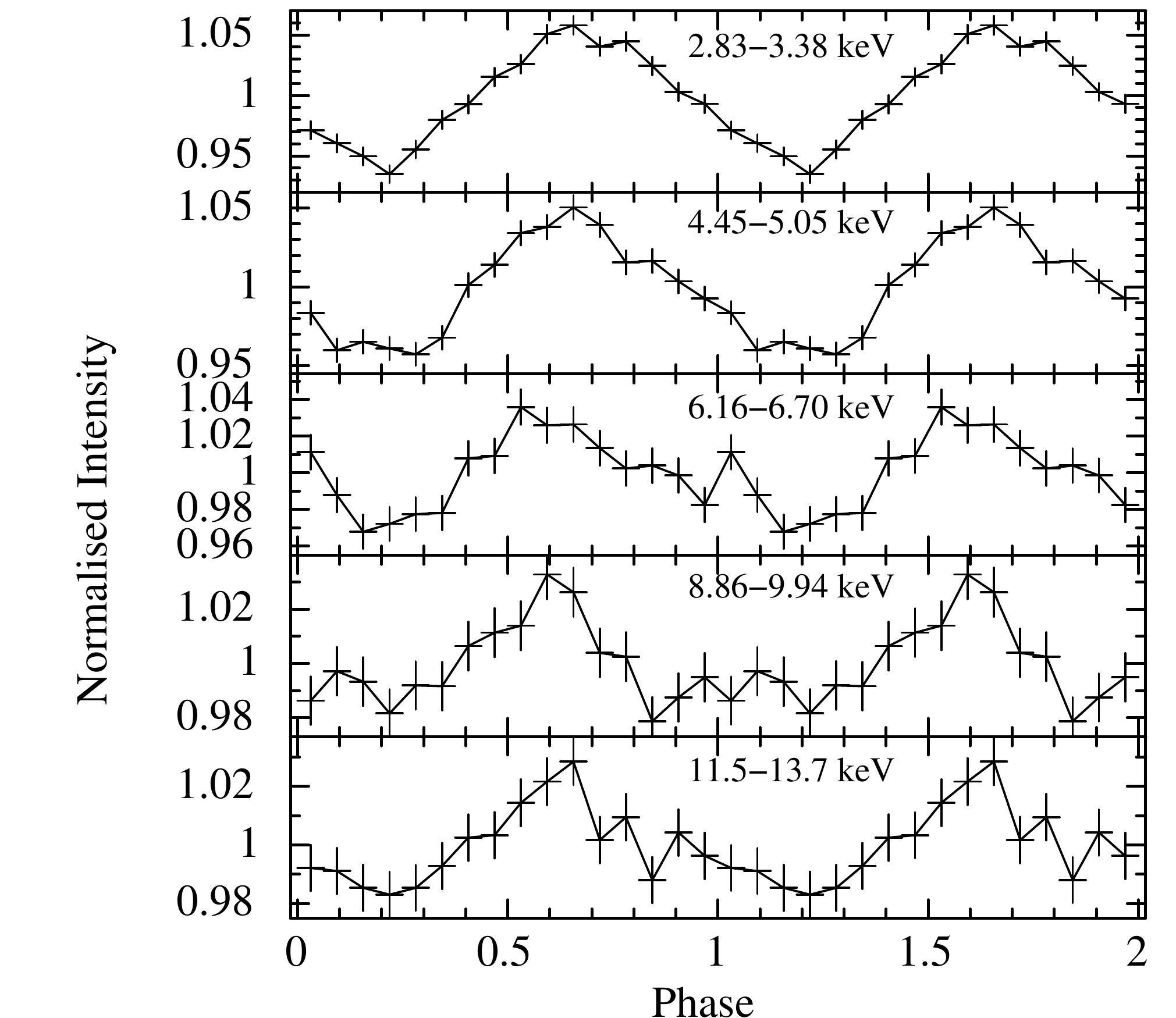}
 \caption{Energy-resolved pulse profiles of \sax. The energy range used is mentioned in respective panels.}
\label{en-profile}
\end{figure}



\bsp	
\label{lastpage}
\end{document}